\documentclass[a4paper,12pt,onecolumn]{scrartcl} %Koma-Skript-Äquivalent zu "article"
\usepackage[a4paper, margin=2.5cm]{geometry}
\usepackage[english]{babel}            %macht deutsche überschriften
\usepackage[latin1]{inputenc}  %man kann Sonderzeiche wie ü,ö usw direkt eingeben
\usepackage{amsmath}           %macht
\usepackage{amsfonts}          %       Mathe
\usepackage{amssymb}           %              mächtiger
\usepackage{graphicx}          %erlaubt Graphiken einzubinden (.eps für dvi und ps sowie .jpg für pdf)
\usepackage{caption}
\usepackage{subfigure}
\usepackage{ae}                %macht schöneres ß
\usepackage{typearea}	         %ermöglicht änderung des Seitenspiegels
\usepackage{lastpage}          %lässt auf die Seienanzahl zugreifen
\usepackage[margin=10pt,font=small,labelfont=bf]{caption} %macht die Bildbeschriftungen richtig
\typearea{16}                  %stellt Seitenspiegel ein
\usepackage{pdfpages}
\usepackage{underscore}
\usepackage{longtable,lscape}
\usepackage{subfigure}
\usepackage{listings}
\usepackage{booktabs}
\usepackage{setspace}
\usepackage{array}
\usepackage{pdfpages}
\usepackage{lmodern}
\usepackage{comment}
\usepackage{tabularx} 
\usepackage{scrlayer-scrpage}
\newpairofpagestyles{firstpage}{
    \clearpairofpagestyles
        \KOMAoptions{footsepline=0.4pt}
    \ifoot{\footnotesize
    \textsuperscript{*}Correspondence: milana.tesfamarian@kaust.edu.sa}
}
\automark{section}
\ihead{}
\ohead{}
\ihead{\rightmark}             % Linke Fußzeile: Section-Name
\ohead{M. Tesfamarian et al.} 
\usepackage{graphicx} % Required for inserting images
\usepackage{graphicx}
\usepackage{placeins}
\usepackage{hyperref}
\usepackage[T1]{fontenc}
\usepackage{float}
\usepackage{hyperref}
\usepackage[T1]{fontenc}
\usepackage{float}

\RedeclareSectionCommand[
  beforeskip=1.2ex plus 0.2ex minus 0.2ex,
  afterskip=0.4ex
]{section}

\RedeclareSectionCommand[
  beforeskip=1.0ex plus 0.2ex minus 0.2ex,
  afterskip=0.3ex
]{subsection}

\RedeclareSectionCommand[
  beforeskip=0.8ex plus 0.2ex minus 0.2ex,
  afterskip=0.2ex
]{subsubsection}

\title{\large \textbf{Numerical Simulation of Transdermal Insulin Delivery Using a Coated Microneedle in a 2D Skin Model}}
  \author{\small
  { Milana Tesfamarian\textsuperscript{1*}, 
 Michael Heisig\textsuperscript{2},
   Gabriel Wittum\textsuperscript{1,2},
 Rolf Krause\textsuperscript{1}}}
\date{\small
  \textsuperscript{1}King Abdullah University of Science and Technology Thuwal, Kingdom of Saudi Arabia \\
  \textsuperscript{2}Goethe University, Kettenhofweg 139, Frankfurt, Germany}
 
\begin{document}

\maketitle
\thispagestyle{firstpage}

\pagenumbering{arabic}
 %\cfoot[Text gerade Seite]{Text ungerade Seite}: Fußzeile Mitte
 %\tableofcontents  % <-- Hier wird das Inhaltsverzeichnis eingefügt
%\newpage
{\centering \section*{\small \centering Abstract}\par}
In this work, we present a computational model to investigate transdermal insulin delivery using coated microneedles. A detailed skin geometry incorporating a coated microneedles was developed to analyze insulin release through the different skin layers and to evaluate the influence of key transport parameters. The model represents the major skin layers: the stratum corneum, viable epidermis, and dermis. Unstructured grids were used to achieve a reliable resolution of the model. The simulations provide insights into the permeation of insulin from the coated microneedles and the transport and distribution across the different skin layers. Finally, the simulation results were compared with experimental data to evaluate the predictive capability of the model.

\newpage
\setlength{\parskip}{1.5em} 
\newpage
\section{Introduction}
\subsection{Insulin Administration and Delivery Strategies}

Since its discovery in 1921, insulin has played a fundamental role in the treatment of diabetes mellitus \cite{gans}. It remains the primary therapy for type 1 diabetes mellitus and is also an important treatment option for patients with type 2 diabetes mellitus. Insulin is mostly administered subcutaneously using different approaches, such as syringes, insulin pens, or continuous subcutaneous insulin infusion systems \cite{shah}. However, subcutaneous insulin administration often requires frequent injections and may be associated with injection-related pain, local skin reactions, and the risk of hypoglycemia \cite{bejal}. To overcome the challenges associated with subcutaneous insulin administration, alternative routes of insulin delivery have been investigated. For instance, pulmonary insulin delivery has been studied as a non-invasive route of administration \cite{elliot}. However, pulmonary insulin delivery is associated with safety concerns, including hypoglycemia, cough, and potential effects on lung function \cite{selam}. Intranasal administration has also gained attention as a non-invasive and potentially painless approach that bypasses the gastrointestinal tract \cite{salzman}. Despite these advantages, intranasal insulin delivery is limited by the low permeability of large molecules across the nasal mucosa and rapid mucociliary clearance, which can result in variable and inconsistent absorption \cite{yaturo}. Overall, significant challenges remain in the development of alternative insulin delivery methods. These challenges emphasize the need for more alternative delivery techniques. Among these approaches, microneedles (MNs) have received increasing attention as a promising alternative for drug administration. In parallel, experimental studies provide essential information on MN-assisted transdermal insulin delivery, but they offer limited insight into the spatial and temporal transport processes occurring within the individual skin layers. Therefore, mathematical and numerical modeling are required to support experimental investigations and provide a quantitative description of insulin transport across the skin.

\subsection{Microneedles for Insulin Administration}
\begin{figure}[H]
\centering
\includegraphics[scale=0.48]{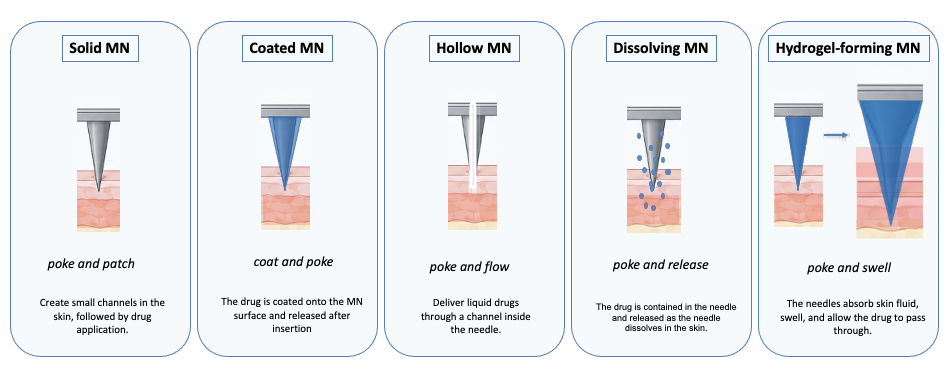}
\caption{Schematic overview of different MN types and their mechanisms for transdermal drug delivery.}
\label{fig:skinMN1}
\end{figure}

Transdermal drug delivery and disease monitoring have gained increasing interest in recent years \cite{hulimane}. Compared with conventional routes of administration, they offer advantages such as improved patient compliance and reduced side effects \cite{nyng,ita}. However, the effectiveness of such approaches is often hindered by the stratum corneum's barrier function. Compounds with a molecular weight (MW) greater than 500 Da have a limited ability to permeate the skin. For example, insulin has a MW of approximately 5.8 kDa, making its transport through the stratum corneum (SC) particularly challenging. MN-based treatments have emerged as a promising strategy to overcome this limitation \cite{qallaf}. By creating microscopic pathways through the SC, MNs enable direct access to deeper skin layers. Moreover, the small size and sharp tips of MNs can reduce tissue damage, pain, and the risk of infection during insertion \cite{kaush, chandra,gill}. Furthermore, in vivo experiments in diabetic rats have demonstrated that transdermally delivered insulin has a consistent pharmacodynamic effect, comparable to that of a subcutaneously administered insulin formulation \cite{ling}. Other in vivo studies have also demonstrated that insulin delivery using MNs can lead to significant and sustained reductions in blood glucose levels \cite{wung}. Various types of MNs (see Figure \ref{fig:skinMN1}) have been developed for different applications and exist in various forms, such as solid, hollow, dissolving, coated, and hydrogel-forming MNs
\cite{mohzin, abian}. 
%Numerical methods provide a valuable framework for evaluating the performance of such delivery systems before fabrication and experimental testing. 

However, the efficient application of MNs requires a much deeper quantitative understanding. Mathematical modeling could help bridge this gap. Among the different types of MNs, this study focuses on the modeling and numerical simulation of  coated microneedles (CMNs) for the transdermal delivery of insulin, in which insulin is deposited as a thin coating onto the surface of solid MNs. During insertion, the coating is delivered directly into the skin, enabling localized and rapid delivery. Despite growing experimental interest in MNs, a detailed mechanistic understanding of insulin transport associated with CMNs remains limited. In particular, the interplay between localized insulin delivery and its subsequent transport through the multilayered skin structure is not yet fully understood. Numerical methods provide a valuable framework for investigating these transport processes before fabrication and experimental testing. In this study, we therefore employ a numerical approach to investigate insulin transport following localized delivery from a CMN within a multilayered skin model.  \\

\section{Mathematical Model}
\begin{figure}[t]%[H]
\centering
\includegraphics[scale=0.47]{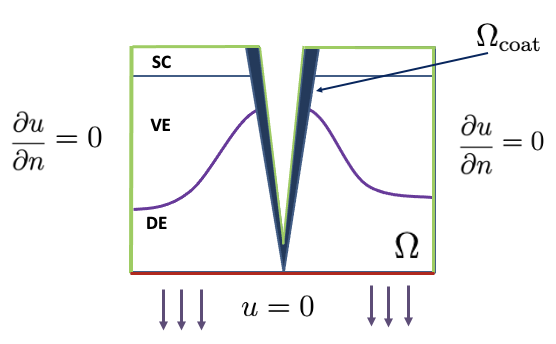} 
\caption{Schematic representation of the two-dimensional computational model with initial and boundary conditions.}
\label{fig:skinMN2}
\end{figure}

We consider a piece of skin $\Omega$ consisting of three layers and a needle
with the coating $\Omega_{\mathrm{coat}}$ (see Figure~\ref{fig:skinMN2}).
A finite dose of insulin is applied to the CMN, allowing the initially loaded
amount to deplete over time. The diffusive transport of insulin is described
by \cite{wittum1}
\begin{equation}
\frac{\partial (K_i u)}{\partial t}
=
\nabla \cdot \left(D_i K_i \nabla u\right),
\qquad
(x,t) \in \Omega_i \times (0,T],
\label{eq:diffusion}
\end{equation}
where $\Omega_i$ denotes the corresponding subdomain and $T$ the end time.
The diffusion and partition coefficients $D_i$ and $K_i$ are assumed to be
constant within each subdomain.

To account for partitioning between adjacent subdomains, a continuous
reference concentration $u(x,t)$ is introduced. The physical insulin
concentration $c_i$ in subdomain $\Omega_i$ is related to the reference
concentration by
\begin{equation}
c_i = K_i u.
\end{equation}
This formulation allows the physical concentration to be discontinuous
between adjacent subdomains while maintaining continuity of the reference
concentration across their interfaces. For
$\Gamma_{ij}=\partial\Omega_i\cap\partial\Omega_j$, conservation of mass
requires continuity of the diffusive flux,
\begin{equation}
D_i K_i \nabla u_i \cdot \mathbf{n}
=
D_j K_j \nabla u_j \cdot \mathbf{n}
\qquad \text{on } \Gamma_{ij},
\label{eq:flux_u}
\end{equation}
where $i$ and $j$ denote neighboring subdomains and $\mathbf{n}$ denotes
the interface normal.

The finite insulin dose in the coating is prescribed by the initial condition
\begin{equation}
u(x,0)=
\begin{cases}
u_0, & x \in \Omega_{\mathrm{coat}},\\
0,   & \text{otherwise}.
\end{cases}
\end{equation}

Since blood capillaries are not explicitly represented in the model, a sink
condition is imposed at the lower boundary of the dermis to represent insulin
removal from the computational domain:
\begin{equation}
u = 0,
\qquad \text{on } \Gamma_{\mathrm{BOT}}.
\end{equation}

The simulations focus on local release, diffusion, and distribution of insulin within the skin rather than explicitly describing vascular uptake and subsequent systemic insulin kinetics. In addition, the mathematical formulation is based on our previously developed finite-dose skin transport model~\cite{me}. 

\section{Numerical methods}
\begin{figure}[H]%[H]
\centering
 \includegraphics[width=0.5\textwidth]{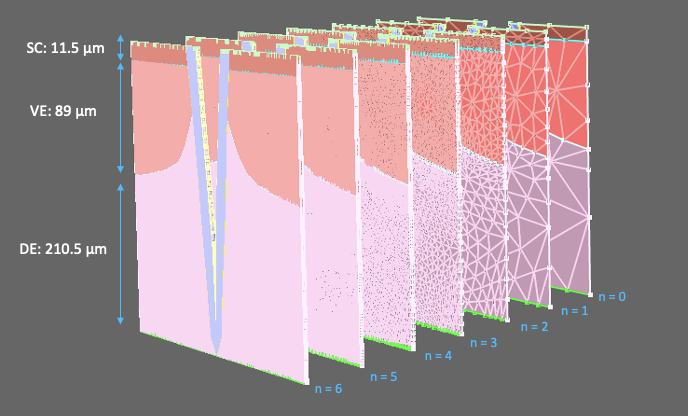}
\caption{2D computational mesh of the CMN-skin geometry with 6 levels of mesh refinement.}
\label{fig:skinMN3}
\end{figure}

The domain $\Omega$ is covered by an unstructured triangular mesh resolving the layer boundaries. The computational grid was uniformly refined over six refinement levels (see Figure \ref{fig:skinMN3}) with up to approximately $2\cdot 10^5$ vertices on the finest grid. The governing equation (\ref{eq:diffusion}) was discretized in space using a vertex-centered finite volume method, ensuring mass conservation across the computational domain, while for the discretization in time we use the LIMEX scheme to capture the time-dependent transport of insulin \cite{uu,ux}. The adaptive time-stepping strategy adjusts the time-step size throughout the simulation, allowing temporal changes in insulin transport to be resolved efficiently. For the solution of the linear systems, we used the BiCGStab method preconditioned with V-cycle, a geometric multigrid (GMG) method. A residual tolerance of $10^{-15}$ was used as the convergence criterion, with BiCGStab converging in under 16 iterations. Furthermore, three ILU pre- and postsmoothing steps were applied at each grid level \cite{hackbuschNeu}.

\section{Model Geometry and Parameters}
\subsection{Model Geometry}
\label{subsec:modelgeometry}
We constructed the computational skin geometry as a two-dimensional layered domain (see Figure \ref{fig:skinMN2}). The SC represents the outermost layer and has a thickness of $11.5\mu\mathrm{m}$ \cite{kano}. Below the SC, the viable epidermis was assigned a thickness of 89.1 $\mu$m, consistent with the experimentally reported range of 50.5 -- 149.1 $\mu$m for human abdominal skin \cite {limcharoen}. The interface between the VE and DE was represented by the characteristic undulating morphology of the dermal--epidermal junction \cite{me,levine}. Below the dermal papillae, the computational domain was extended to capture the deeper dermal region. The MN had a total length of approximately $222\mu\mathrm{m}$ and was modeled as fully inserted into the skin \cite{resnik,he}. The base diameter of the MN is 58 $\mu$m, comparable to representative CMN designs reported in the literature \cite{neil}.We modeled the MN as a single needle with the insulin-loaded coating represented as a separate computational subdomain along the MN surface. The coating thickness was approximately 9 -- 11$\mu$m \cite{dana}, while local variations occurred in the immediate tip region as a consequence of the geometric construction.

\subsection{Model Parameters}
\label{sec:model_parameters}
The model parameters were selected based on literature data and a systematic parameter analysis. Due to the limited availability of numerical data for insulin, experimentally reported values for insulin in skin tissue were considered. Layer-specific diffusion coefficients were assigned to account for differences in molecular transport across the skin, with lower values in barrier-dominant regions and higher values in the underlying viable skin layers. For the CMN, a low diffusion coefficient of $D_{\mathrm{coat}} = 0.6~\mu\mathrm{m}^2/\mathrm{min}$ was selected to represent the slow release of insulin \cite{leela}. Higher diffusion coefficients led to a substantially faster depletion of the coating and more rapid insulin release into the surrounding tissue. Since experimental data for insulin partition coefficients in the individual skin layers are limited, the partition coefficients were systematically varied to investigate their influence on insulin transport, while the diffusion coefficients were kept constant. A detailed analysis of this parameter study will be presented in a future publication. The selected diffusion and partition coefficients reflect the expected differences in insulin transport across the individual skin layers and follow the hierarchy
\begin{equation}
K_{\mathrm{SC}} \ll K_{\mathrm{VE}} \ll K_{\mathrm{DE}},
\qquad
D_{\mathrm{SC}} \ll D_{\mathrm{VE}} \ll D_{\mathrm{DE}}.
\end{equation}
This hierarchy reflects the low affinity and restricted diffusion of insulin in the SC and the higher partitioning and diffusion in the viable skin layers.

\section{Results}\label{sec:rdiscussion}
The numerical simulations were performed using  UG4 \cite{ug41,ug42}, a software that provides efficient numerical methods for solving partial differential equations on unstructured grids in one, two, and three dimensions. The two-dimensional computational geometry and corresponding grid were implemented using the meshing software ProMesh \cite{proM}.
In the following, experimental and simulation data will be compared. We evaluated time-dependent changes in dermal uptake over 500 min. To assess the influence of the skin morphology and CMN geometry on transport behavior, we analyzed several simulation cases with different epidermal thicknesses and varying channel depths $d_c$ and widths $w_c$. The configuration introduced in Section~\ref{subsec:modelgeometry} was used for the comparison with experimental data. Future work will include further parameter studies and additional comparisons with experimental data.

\subsection{Evaluation of the Numerical Model Using Experimental Data}\label{subsec:experiment}

\begin{figure}[H]
\begin{center}
\includegraphics[scale=0.47]{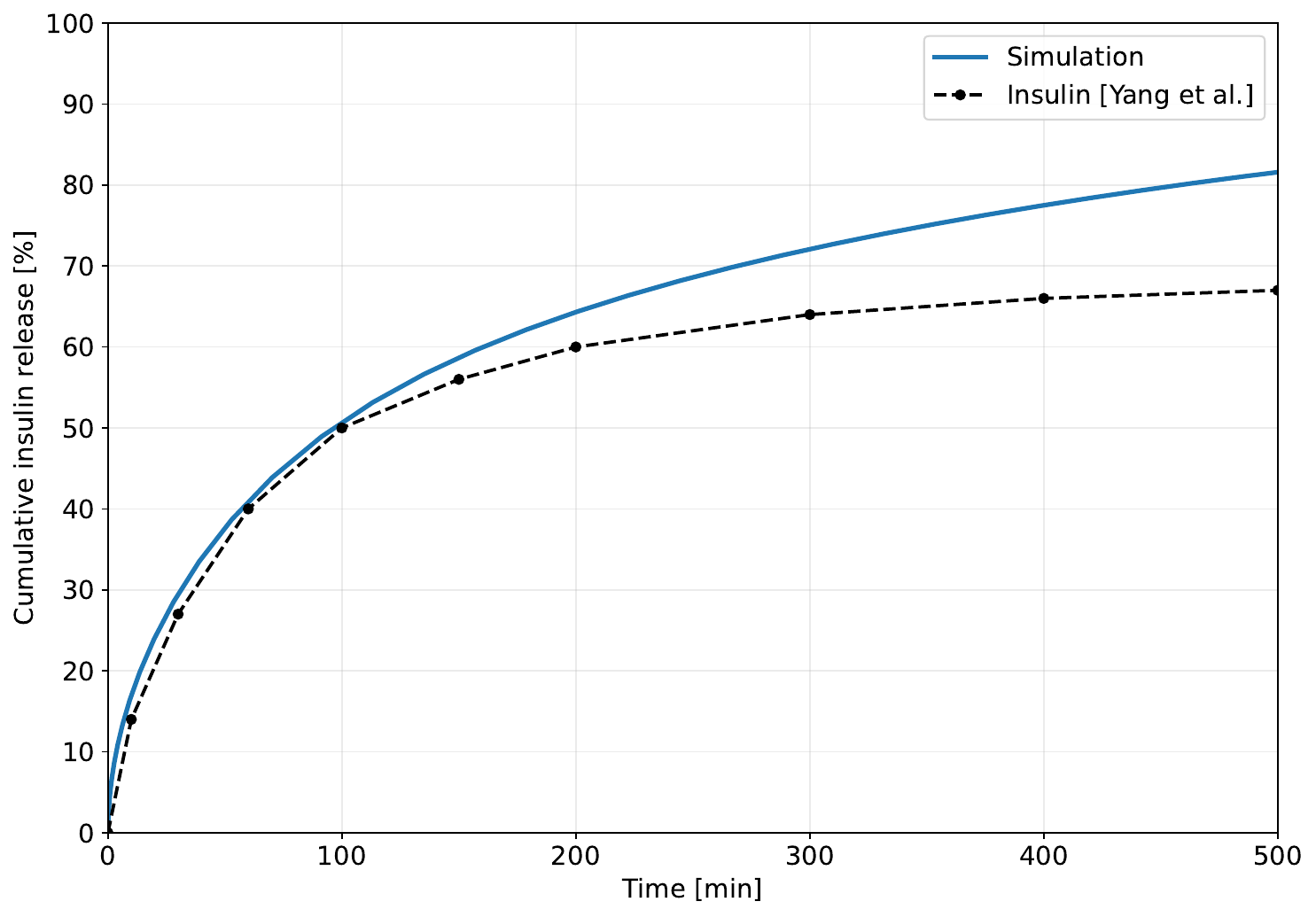}
\end{center}
\caption{ Comparison of simulated cumulative insulin permeation with experimental data reported by Yang et al. \cite{yangp}.}
\label{fig:insurelease}
\end{figure}

The simulated release profile of the CMN was compared with the numerical results reported for hydrogel MNs \cite{yangp}. In the hydrogel MN study, a coupled swelling--obstruction--mechanics model was employed to describe drug release while accounting for hydrogel swelling, molecular transport, and the mechanical interaction between the MNs and the skin. The model was further applied to different drug molecules: dextrans with molecular weights of 4, 10, and 20 kDa, and insulin (5.8 kDa), which exhibited slower release than the other drug profiles. Figure \ref{fig:insurelease} compares the cumulative insulin release simulated by our CMN model and reported by Yang et al. for hydrogel MNs. During the initial release phase, both profiles show similar behavior, with a rapid increase in cumulative insulin release. Up to approximately 100 min, the simulated CMN profile is consistent with the experimental hydrogel MN data. At this time point, both profiles show a cumulative insulin release of approximately 50\%. However, after around 120 min, the profiles gradually diverge. The hydrogel MN release approaches a plateau at approximately 67\% after 500 min, whereas the CMN simulation continues to increase and reaches approximately 81\% over the same period.

The differences at later time points could be due to the distinct release mechanisms of the MN. In hydrogel MNs, insulin release is influenced by the swelling of the polymer matrix and diffusion through the swollen hydrogel network, whereas in the CMN model, insulin is released from a thin coating directly into the surrounding skin tissue. Differences in MN geometry and dimensions, including MN length, as well as material properties and transport parameters, may further contribute to the deviation between the profiles. Nevertheless, the similar behavior during the initial release phase shows that the CMN model reproduces a temporal release pattern comparable to that reported for hydrogel MNs.

\newpage
%\clearpage
\section{Conclusion}\label{conclusion}
This study presented a numerical model for investigating insulin transport with the application of a single CMN. The model represented different skin layers, allowing analysis of insulin transport through the multilayered skin structure. The simulations present the time-dependent transport of insulin from the CMN region into the surrounding tissue and its subsequent permeation through the skin. The numerical results were further compared with experimental insulin release data obtained using hydrogel MNs. The comparison showed a similar overall trend in cumulative insulin release. Overall, the results demonstrate the applicability of the numerical approach for investigating insulin transport using CMN-based delivery. Future work will include additional parameter studies, alternative CMN representations, and further comparisons with experimental data to improve and evaluate the model.

\section*{Acknowledgments}
This research was supported by King Abdullah University of Science and Technology (KAUST).

\section*{Conflict of interest statement}
The authors declare that they have no known competing financial interests or personal relationships that could have appeared to influence the work reported in this paper.

\section*{Data availability statement}
All data supporting the findings of this study will be available in the published article.

\pagestyle{plain}

\end{document}